Effect of Feedback between Environment and Finite Population

**Authors:** Jia-Xu Han[1,2], Rui-Wu Wang[1]*

1. School of Ecology and Environment, Northwestern Polytechnical University, Xi'an 710072, PR China
2. Zoology Department and Biodiversity Research Centre, University of British Columbia, Vancouver, British Columbia V6T 1Z4, Canada

*Corresponding author: Rui-Wu Wang (wangrw@nwpu.edu.cn)

**Author contributions:** J-X. H. and R-W. W. contribute to modeling and write the paper together.

**Acknowledgments:** Discussion with Dolph Schluter and Feng Zhang improved the models. This research was supported by National Natural Science Foundation of China-Yunnan Joint Fund (U2102221).

# Abstract

Natural selection imply that any organisms including human being will evolve to improve its fitness advantage and the selected genotype or phenotype in equilibrium state will not vary over the time. However, evolutionary process of biological organisms in reality is greatly affected by the environmental change and historical accidents. In this research, we construct a co-evolutionary system to investigate the impact of species-environment feedback. When we talk about an invasion species or mutation, positive feedback is detrimental to the success of the invasion because positive feedback benefits a large number of individuals, whereas negative feedback benefits the invasion because negative feedback disadvantages a large number of individuals. In the case of a competition between two species with initially equal numbers of individuals, both positive and negative feedback will favor the species with low fitness, increasing its chances of taking over the whole population. The reason for this is that feedback allows initially inferior species to have greater fitness than initially dominating species in the early stages, emphasizing the importance of early random accident. Our findings emphasize the significance of the evolutionary path driven by species-environment feedback.

# 1. Introduction

Natural selection of Darwin implies that biological organisms are evolved gradually in a direction to increase fitness due to selfness of organisms, which is an idea of rationality of human being proposed by Adam Smith (Darwin 1859, Burkhardt & Smith 1985). In this theoretical framework, the individuals or the populations will reach to equilibrium and therefore predictable through the competition among each other, especially when the shared resource is saturated, and it is also the fundamental concept of Modern Synthesis in theory of evolutionary biology (Huxley 1942; Wang 2023). Competition as the mechanism for the nature selection pushes evolution towards optimum fitness, which also constitutes a balance. Ecology and evolution research nowadays frequently highlights this balanced perspective of dynamic systems by resolving a model that identifies and defines the stability of their balances (Spencer 2020). The mathematical ecologists generally ignore history accidents since they believed it was smacked of particularities and had little generality (Kingsland 1985).

However, the historical accidents might play an important role in evolutionary process that even might change the species speciation or population and ecosystem structure such as priority effect. The first species to arrive in a habitat alters the resources available to other species, making the environment more or less appropriate for them (Jared 1975; Drake 1991). In the theoretical exploration, historical accidents described by the evolutionary path (Blount et al. 2018), are treated as a feedback process of adaptation in which the organism has an effect on its own development as both the subject of the natural selection and its producer (Laland et al. 1999;

Wang et al. 2021). For example, the organisms can change the environment by their metabolism, activities and choices (Odling-Smee et al. 1996; Nowak & Sigmund 2004). A change at any point in the sequence of historical accidents can change the environment thereafter, and thus affect the natural selection that organisms face afterwards. These adaptation feedback in environment may functions as negative feedback, which usually lead to equilibrium and predictable results, or positive feedback, which allows for evolution to adapt faster than evolution by selection from the unchanging environment (Odling-Smee et al. 2003, Laland et al. 2016). The positive feedback causes every stage of development depends on its preceding phases (Arthur 1989). Then, historical events may affect the evolutionary path and the sequence of history has a major impact on the result (e.g. Blount et al. 2008, Zheng et al. 2020).

In evolutionary biology, stochasticity in the evolutionary process is integral part of much of population genetic theory and phylogenetic modelling when the population size is finite. One phenotype is frequently discovered to be marginally beneficial over another phenotype and one of these phenotypes may be selected according to stochastic process (Charlesworth & Eyre-Walker 2007). Even traits that are not favored by natural selection have the ability to take over a whole population, which is a key distinction between deterministic and stochastic models of evolution (Nowak 2006). However, there has been little research into the stochasticity of a finite population in the setting of feedback with the environment. To emphasize the relevance of historical accidents, we employ a finite population model to explore the influence of feedback with the environment on the probability of a species taking over the whole population. To begin, we construct a co-evolutionary process between species and environment. Then, in the absence of feedback, we compute the analytic solution for the finite population dynamics. Finally, we incorporate feedback and demonstrate how it works using individual-based simulation.

## 2. Model

In this section, we look at how species and their environments co-evolve. The term "species" in this sense includes not only species abundance but also gene frequency or other relevant features. It might also be a trait or a mutation. The term "environment" might refer to a resource that benefits the species, and we will use "resource" instead of "environment" later in the text to simplify the explanation. And whether the resource is created by the species or just used by the species, there will be positive or negative feedback between the species and the environment. To keep things simple, we disregard the influence of density dependence on species and resource dynamics. As an example, consider two species with two resources.

First, we explore the role of the environment in the co-evolutionary process. We assume that there are no relationships between the resources, allowing us to study their dynamics individually. This might be a microcosm of each species' own resource, while they share certain resources. We propose that the dynamic of the resource itself obeys exponential growth, similar to the generalized Lotka-Volterra equation, which is a frequently used tool for understanding the dynamics of interacting populations (Wangersky 1978). And, the more contacts there are or the more resources there are, the faster the resources and species are eaten, and the more suitable it is for that resource to increase. The resource dynamic may thus be expressed as:

$$\frac{dY_i}{dt} = Y_i(a_i X_i + b_i), \quad i = 1,2,$$

where $Y_i$ represents the amount of resource $i$, $X_i$ represents the number of individuals of the species $i$ that is paired with resource $i$, $a_i$ represents the influence of species presence on resource growth rate, and $b_i$ represents the highest per unit resource growth rate. To indicate the ratio of total resources for resource $i$, we may use $y_i = \frac{Y_i}{Y_1 + Y_2}$. The dynamic of $y_i$ then follows:

$$(Y_1 + Y_2)\frac{dy_i}{dt} = \frac{d(Y_1+Y_2)y_i}{dt} - \frac{d(Y_1+Y_2)}{dt} y_i = \frac{dY_i}{dt} - \frac{d(Y_1+Y_2)}{dt} y_i$$

$$= Y_i(a_i X_i + b_i) - y_i \sum_{j=1}^{2} Y_j(a_j X_j + b_j).$$

Then, we can rewrite the dynamic of $y_i$ as

$$\frac{dy_i}{dt} = y_i(a_i X_i + b_i) - y_i \sum_{j=1}^{2} y_j(a_j X_j + b_j)$$

$$= y_i(1 - y_i)(a_i X_i + b_i - a_{3-i} X_{3-i} + b_{3-i}).$$

Because we are interested in the effect of feedback, we simplify the model by assuming that the values of the parameters for these resources are equivalent, i.e., $a_1 = a_2 = a$ and $b_1 = b_2$. We can also get

$$\frac{dy_i}{dt} = a y_i(1 - y_i)(X_i - X_{3-i}). \quad (1)$$

The parameter $a$ can be used to quantify the qualities of feedback between environment and species, with $a > 0$ indicating positive feedback, $a = 0$ indicating no feedback, and $a < 0$ indicating negative feedback. And the $|a|$ can quantify the intensity of feedback between the environment and the species.

We assume that the overall population size is fixed and finite in each step of the process, and that only one new member is produced and one member is chosen to die (birth-death process) in each time step, which is a general stochastic process based on Moran's model in the presence of demographic stochasticity (Moran 1958). The new member is assigned the same phenotype as one randomly selected existing member, with a probability equal to the proportion of that member's fitness divided by the total fitness of all members. And the member chosen to die is picked at random, with a uniform distribution across the species. Because the union resource $i$ will only be used by species $i$, the advantages received by species $i$ are $f_i = c_i y_i + d_i$. Because the species gain from the shared resources, the union resource only makes a minor contribution to fitness, which is emphasised by weak selection. Weak selection describes situations in which reward differences have minimal influence, allowing random fluctuations to dominate evolutionary dynamics (Wild & Traulsen 2007, Taylor et al. 2007). The fitness of species $i$ will be $F_i = 1 - w + w f_i$, where $w$ indicates the intensity of selection.

In this context, we evaluate the probability $\rho_k$ of a species $i = 1$ with $k$ individuals taking over a population with a total population of $N$ individuals. A similar scenario exists for the species $i = 2$. Then we look at two scenarios. To begin, we evaluate the fix probability $\rho_1$ of species 1, which is the probability that species 1 will ultimately take over the whole population when the population initially comprises just one individual from species 1 and all other individuals are from species 2. This condition is analogous to when a species accidentally invades another species' environment or when a mutant develops by chance in a population (invasion scenario). The fix probability will then be the probability that the invader or mutation will take over the whole population. The alternative conditions is $\rho_{\frac{N}{2}}$, which is the probability that species 1 will ultimately take over the whole population if the population begins with half of the individuals from species 1 and half of the populace from species 2. This circumstance relates to a fair rivalry between the two species, such as the dynamics at the intersection of the distribution zones of two species (competition scenario). In this fair fight, the probability $\rho_{\frac{N}{2}}$ denotes the probability that species 1 will defeat species 2.

## 3. The situation without feedback

To begin, we investigate the case when there is no feedback between species and environment, which is $a = 0$, as a baseline for future outcomes. The co-evolution system will then degenerate into a simple finite population process, and we will set $y_1 = y_2 = 1/2$. In each phase of the Moran process, one individual is randomly picked to produce offspring and one individual is randomly selected to die. The possibility of species 1 with *k* individuals increasing to *k+1* individuals is the probability of an individual of species 1 chosen to birth multiplied by an individual of species 2 chosen to die, and it may be represented as

$$p_k^+ = \frac{kF_1}{kF_1 + (N-k)F_2} \frac{N-k}{N}.$$

The possibility of species 1 with *k* individuals drops to *k-1* individuals is the probability that an individual of species 2 chosen to be born times an individual of species 1 chosen to die, and it may be represented as

$$p_k^- = \frac{(N-k)F_2}{kF_1 + (N-k)F_2} \frac{k}{N}.$$

Except for the absorbing states, $\rho_0 = 0$ and $\rho_N = 1$, the dynamic of this finite population will follow

$$\rho_k = \rho_{k+1} p_k^- + \rho_k (1 - p_k^- - p_k^+) + \rho_{k-1} p_k^+,$$

because the condition of *k* people originates from three circumstances: *k+1* individuals reduce one individual, *k* individuals remain stable, and *k-1* individuals grow one individual. To solve this equation, we follow Nowak (2006) and Ohtsuki et al (2007). Let $\alpha_k = \rho_k - \rho_{k-1}$ and the dynamic of this finite population be written as $\alpha_k = \frac{p_k^-}{p_k^+} \alpha_{k-1}$. Because $\sum_{k=1}^N \alpha_k = \rho_N - \rho_1 = 1$ and $\alpha_k = \prod_{j=1}^{k-1} \frac{p_j^-}{p_j^+} \rho_1$, we may have $\rho_k = \frac{1 + \sum_{j=1}^{k-1} \prod_{l=1}^j \frac{p_j^-}{p_j^+}}{1 + \sum_{j=1}^{N-1} \prod_{l=1}^j \frac{p_j^-}{p_j^+}}.$

Because the selection is weak ($w \ll 1$) in the invasion scenario, we may calculate an approximation of the fix probability,

$$\rho_1 = \frac{1}{1 + \sum_{j=1}^{N-1} \prod_{l=1}^j \frac{p_j^-}{p_j^+}} = \frac{1}{1 + \sum_{j=1}^{N-1} \prod_{l=1}^j \frac{F_2}{F_1}} = \frac{1}{1 + \sum_{j=1}^{N-1} \prod_{l=1}^j \frac{1 - w + w f_2}{1 - w + w f_1}}$$

$$= \frac{1}{1 + \sum_{j=1}^{N-1} \prod_{l=1}^j \left(1 + w \frac{f_2 - f_1}{1 - w + w f_1}\right)} \approx \frac{1}{1 + \sum_{j=1}^{N-1} \prod_{l=1}^j [1 + w(f_2 - f_1)]}$$

$$\approx \frac{1}{1 + \sum_{j=1}^{N-1} [1 + \sum_{l=1}^j w(f_2 - f_1)]} = \frac{1}{N + w \frac{N(N-1)}{2}(f_2 - f_1)}$$

$$= \frac{1}{N} \frac{1}{1 + w \frac{(N-1)}{2}\left(\frac{c_2}{2} + d_2 - \frac{c_1}{2} - d_1\right)}.$$

If the process is completely random, which implies that all of the parameters of species 1 and species 2 are the same, the probability of species 1 originally containing k individuals taking over the population is $\frac{k}{N}$. Then, if $\rho_1 > \frac{1}{N}$, we may argue that natural selection favors species 1, which is equal to

$$f_2 = \frac{c_2}{2} + d_2 < \frac{c_1}{2} + d_1 = f_1. \qquad (2)$$

Because the environment remains constant, it is simple to grasp that natural selection favors species only if the fitness of species 1 is greater than that of species 2. In the competition situation, we get $\rho_{\frac{N}{2}} \approx \frac{1}{2} \frac{1+w\frac{(N/2-1)}{2}\left(\frac{c_2}{2}+d_2-\frac{c_1}{2}-d_1\right)}{1+w\frac{(N-1)}{2}\left(\frac{c_2}{2}+d_2-\frac{c_1}{2}-d_1\right)}$. In this case, $\rho_{\frac{N}{2}} > 1/2$, which is also identical to inequality (2), suggests that species 1 has a better probability of dominating the population than species 2. To replicate this basic Moran process, we also utilize an individual-based model (Figure 1). In the zone satisfying inequality (2), the values of $\rho_1 - \frac{1}{N}$ and $\rho_{\frac{N}{2}} - \frac{1}{2}$ are positive. Then, we refer to species 1 as the initially dominant species if $\frac{c_1}{2} + d_1 > \frac{c_2}{2} + d_2$ is fulfilled, initially inferior species if $\frac{c_2}{2} + d_2 > \frac{c_1}{2} + d_1$ is satisfied, and initially neutral species if $\frac{c_2}{2} + d_2 = \frac{c_1}{2} + d_1$ is satisfied.

Figure 1.

a.

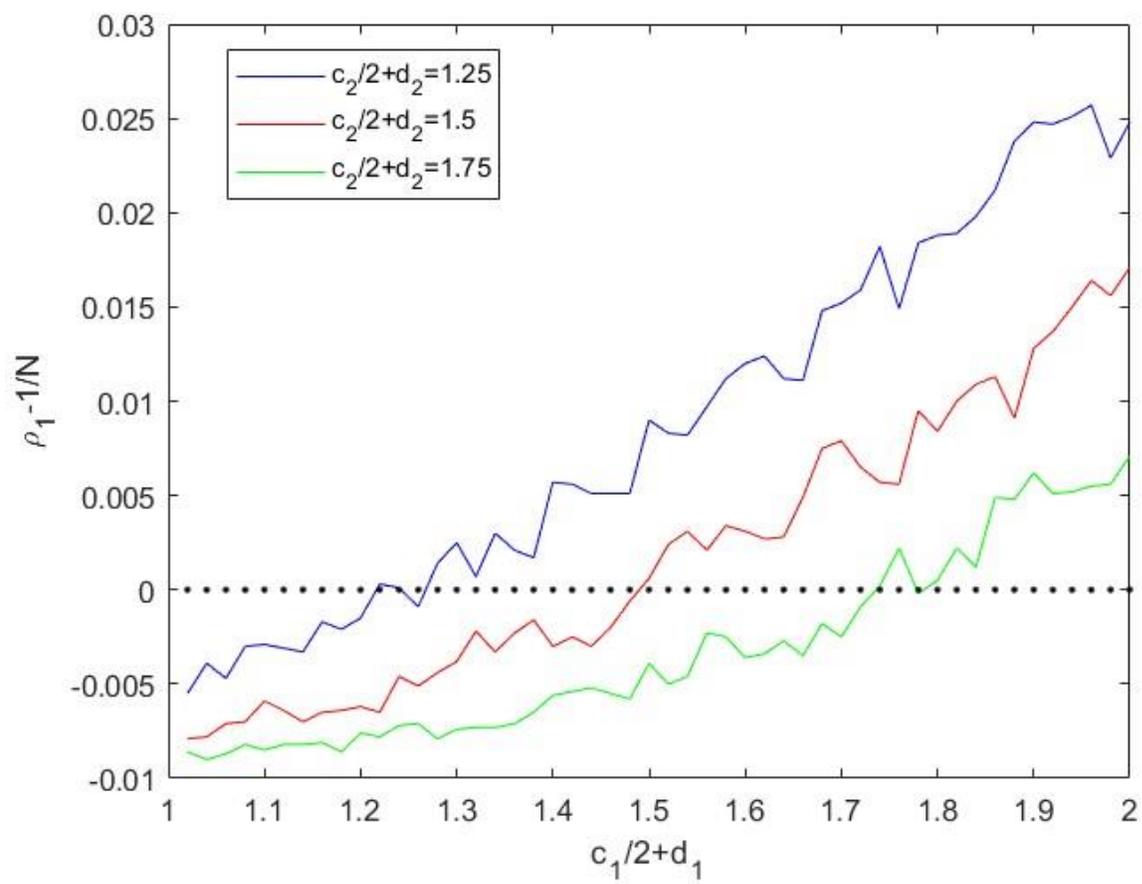

b.

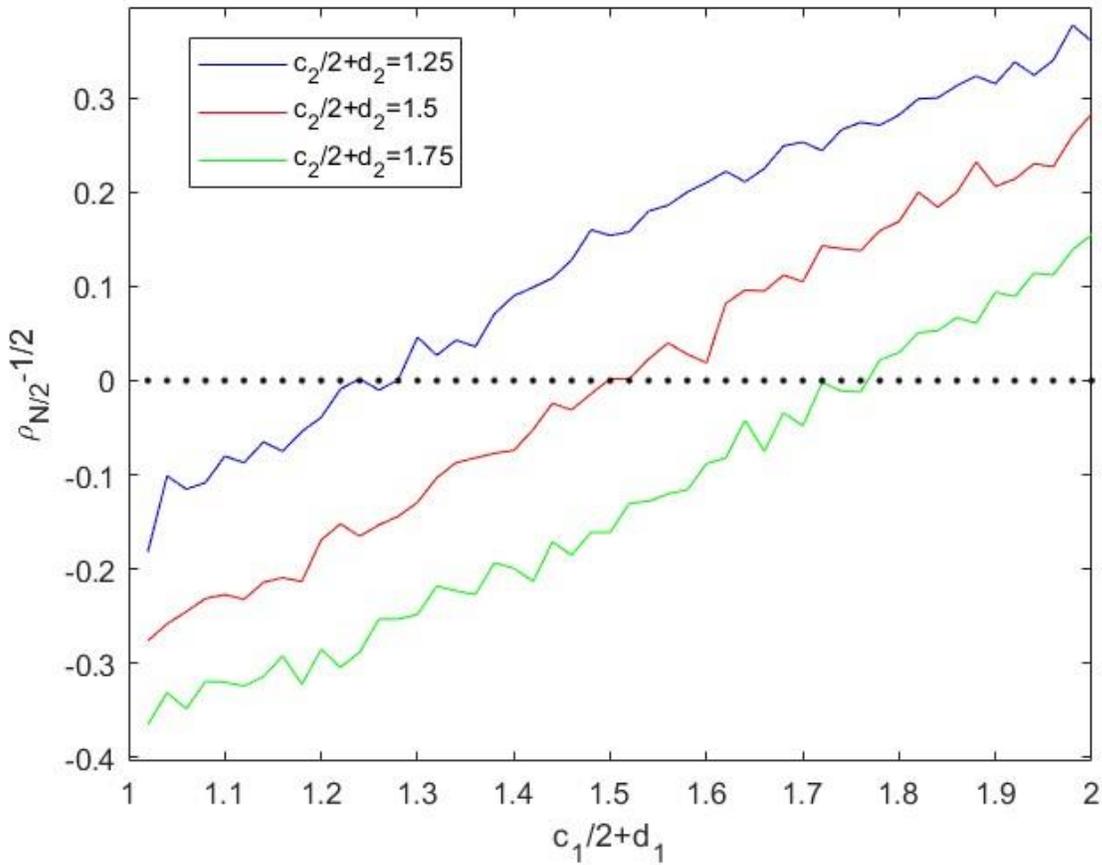

Figure 1. Simply finite population dynamic without feedback between environment and species. If inequality (2) is met, the probability of species 1 taking over the whole population will be greater than the neutral scenario. **a.** One individual from species 1 is present in the beginning population; alternatively, **b.** half of the initial population is from species 1. The entire population size is 100, and the selection intensity is 0.05. Each simulation is conducted a total of 10000 times to compute the probabilities $\rho_{\frac{1}{2}}$ and $\rho_{\frac{1}{N}}$..

## 4. The situation with feedback

When the presence of a species has an effect on the environment, which means $b \neq 0$, stochastic processes in finite populations then have an impact on the environment. Changes in the environment will then alter the fitness of the species and, as a result, the stochastic processes of the finite population. Because of the feedback, this co-evolution system has become so complicated that an analytical solution cannot be provided. And we will continue to analyze this system utilizing individual-based simulation. To simulate, we must rewrite the resource dynamic (equation (1)) as:

$$\Delta y_i = a y_i (1 - y_i)(X_i - X_{3-i})\Delta t,$$

where $\Delta t$ is the time step for modelling differential equations and $\Delta y_i$ represents the resource change per time step. $\Delta t$ is also the time scale between the dynamics of species and environment in this case. And we set the beginning resources for both species to be averaged. The simulations on the species component are consistent with the preceding section, and we assume that the change of species occurs first, followed by the change of environment in one time step. And we set $\frac{c_2}{2} + d_2 = 1.5$ and use $\frac{c_1}{2} + d_1 = 1.75$, $\frac{c_1}{2} + d_1 = 1.5$, and $\frac{c_1}{2} + d_1 = 1.25$ to represent the condition in which species 1 is initially dominating, neutral, or inferior.

4.1 The situation with positive feedback

Positive feedback increases the importance of individual numbers, therefore having a high individual number is advantageous. In the invasion scenario, the fix probability $\rho_1$ declines fast to near 0 as the amount of feedback increases, regardless of whether species 1 is initially dominating, neutral, or inferior (Figure 2a). In other words, positive feedback makes the invasion more difficult than in the absence of feedback, or perhaps impossible. The possibility of the dominant species taking over the population is very low, therefore the initially dominating species that would be favored by natural selection in the absence of feedback are likewise not favored anymore. This occurs because positive feedback permits the species with initial dominance to modify the environment more quickly, making it more appropriate for this species. The improved environment causes the species to be more fit and has a greater chance of resisting invasive success. To be more explicit, examine the first two-time steps of this process for $\frac{c_1}{2} + d_1 = 1.75$. If species 1 is chosen to generate offspring and species 2 is chosen to have a deceased individual in the first step, individuals in species 2 will have a better environment than individuals in species 1 in the following step, which should be the same if no feedback is provided. When compared to the situation without feedback, positive feedback reduces the possibility of species 1 taking over the population. Furthermore, the occurrence of other events in the first step has no impact on the probability of species 1 taking over the population.

Figure 2

a.

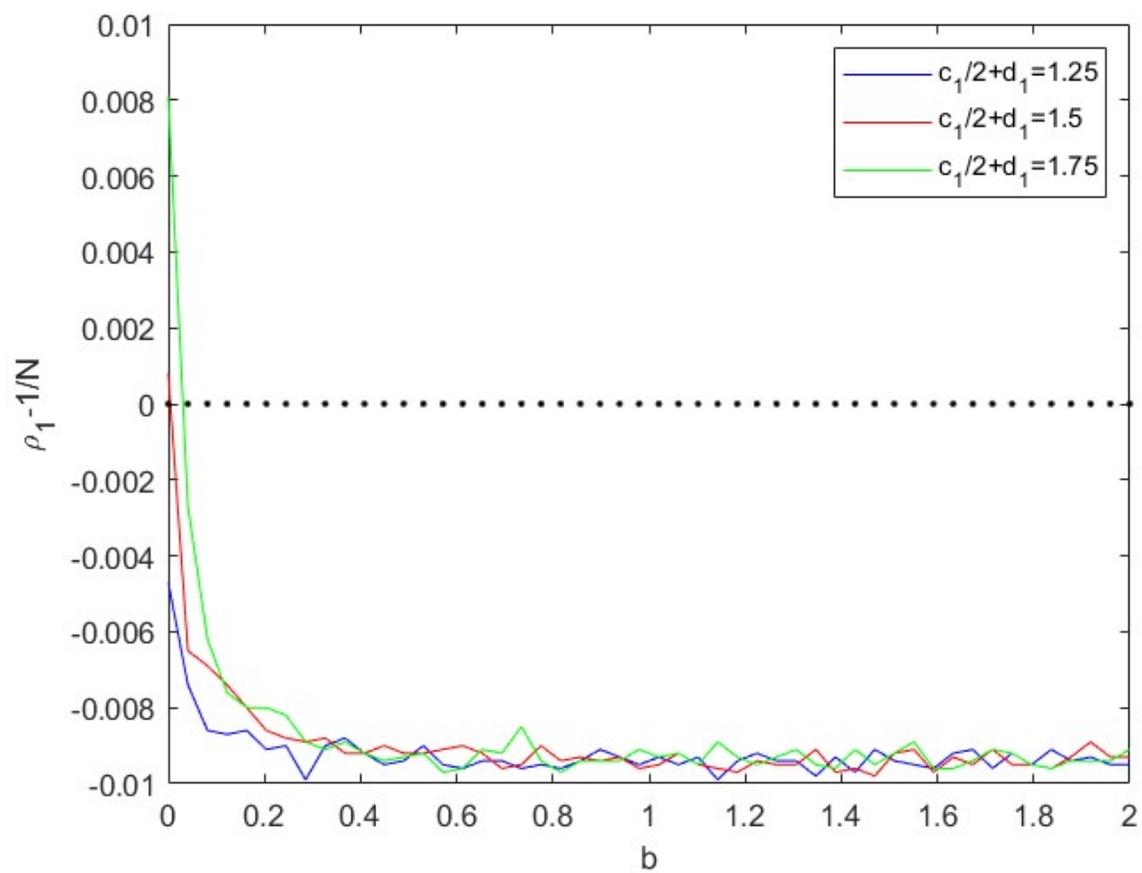

b.

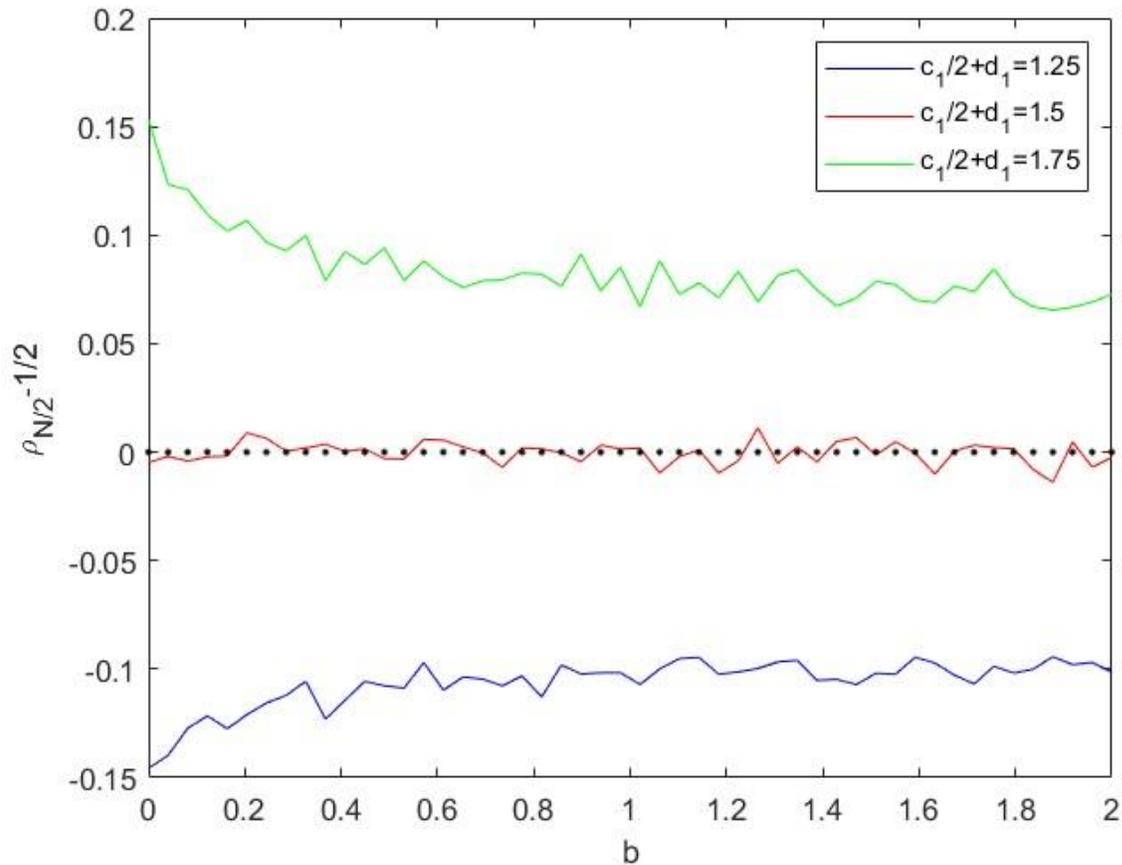

Figure 2. The finite population dynamic is characterized by positive feedback between the environment and the species. **a.** In the invasion scenario, positive feedback makes it difficult for species 1 to invade species 2, regardless of whether species 1 is initially dominating, neutral, or inferior. **b.** In a competing scenario, positive feedback may weaken the initially dominant species while strengthening the initially inferior species. The entire population size is 100, and the selection intensity is 0.05. Each simulation is conducted a total of 10000 times to compute the probabilities $\rho_{\frac{1}{2}}$ and $\rho_{\frac{1}{N}}$. Other parameter: $\Delta t = 0.001$ and $\frac{c_2}{2} + d_2 = 1.5$.

With increasing feedback intensity, the possibility of taking over the whole population decreases for the initially dominant species but increases for the initially inferior species in the competing scenario (Figure 2b). In other words, positive feedback diminishes the dominant species' dominance, making the inferior species more likely to take over the whole population than in the absence of feedback. Positive feedback, however, does not render species favored by natural selection unfavorable and vice versa. Because of demographic stochasticity, the originally inferior species may have more individuals, allowing it to have more resources and hence better fitness. More precisely, positive feedback causes species 1 to have a greater fitness than species 2 in the early stages if species 1 is initially inferior (Figure 3a). Positive feedback makes early stochastic evolutionary paths so significant that they can affect the property of the evolution process. Similarly, positive feedback causes species 1 to lose its advantage if it is the dominant species at first (Figure 3b).

Figure 3

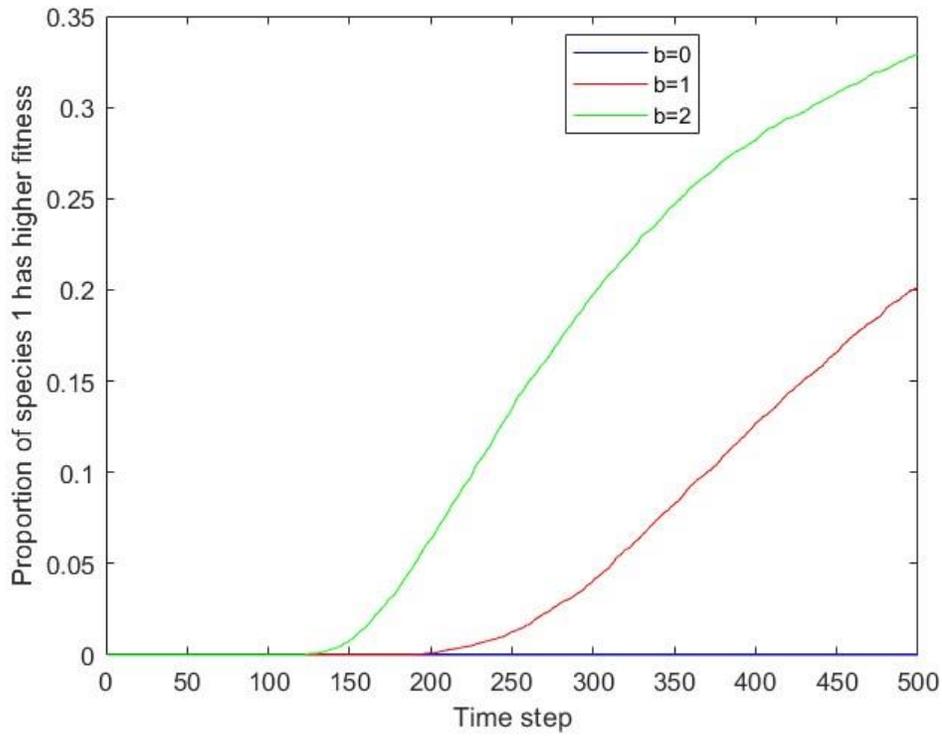

a.

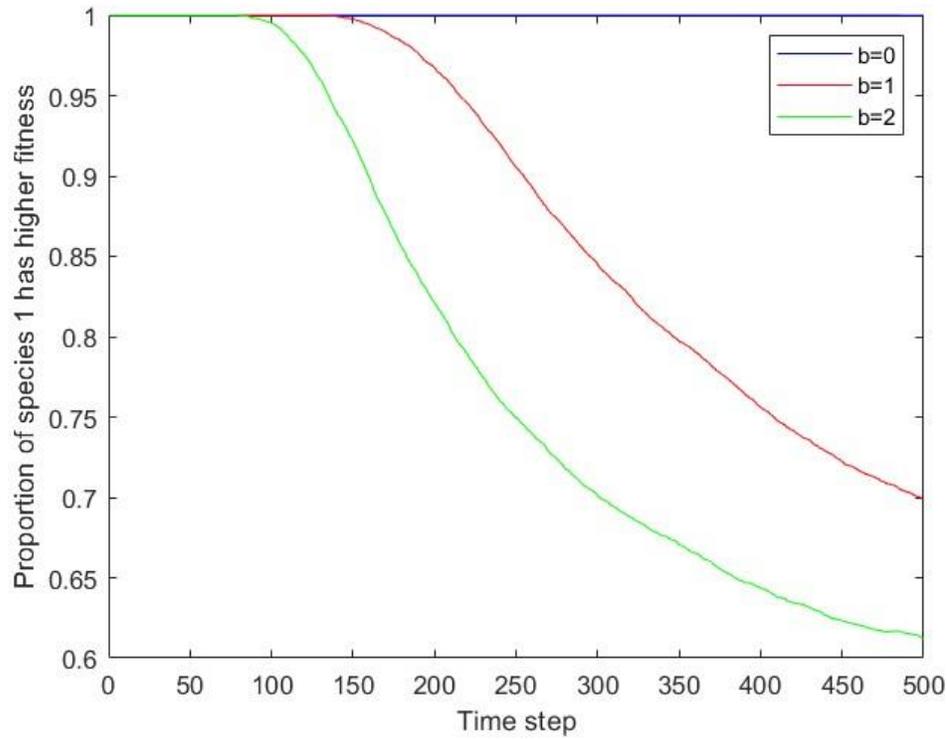

b.

Figure 3. The early effects of positive feedback on fitness. Positive feedback allows **a.** initially inferior species with $\frac{c_1}{2} + d_1 = 1.25$ to gain fitness and **b.** dominant species with $\frac{c_1}{2} + d_1 = 1.75$ to lose its advantage in the first 500 time steps. The entire population size is 100, and the selection intensity is 0.05. Each simulation is conducted a total of 10000 times. Other parameter: $\Delta t = 0.001$ and $\frac{c_2}{2} + d_2 = 1.5$.

4.2 The situation with negative feedback

Individual numbers become disadvantageous as a result of negative feedback. In the invasion scenario, negative feedback has the opposite effect on the outcomes as positive feedback. The fix probability $\rho_1$ increases with increasing feedback intensity, regardless of whether species 1 is initially dominant, neutral, or inferior (Figure 4a). In particular, the fix probability of initially inferior species may be more than $1/N$, which is the fix probability of neutral species in the absence of feedback. The rationale for this is that species with more starting individuals will consume resources more quickly, lowering the fitness of this species. The only individual from species 1 then obtains an advantage.

In a competition scenario, negative feedback serves the same effect as positive feedback. With increasing feedback intensity, the chance of taking over the whole population decreases for the initially dominant species but increases for the initially inferior species (Figure 4b). Similarly, negative feedback has no impact on whether the species is favoured by natural selection. In other words, the possibility of an initially dominant species gaining control of the entire population will not be less than 1/2, and the probability of an initially inferior species gaining control of the entire population will not be more than 1/2. Negative feedback, like positive feedback, causes initially inferior species to have a high probability of fitness and initially dominating species to have a low fitness (Figure 4 c,d).

Figure 4

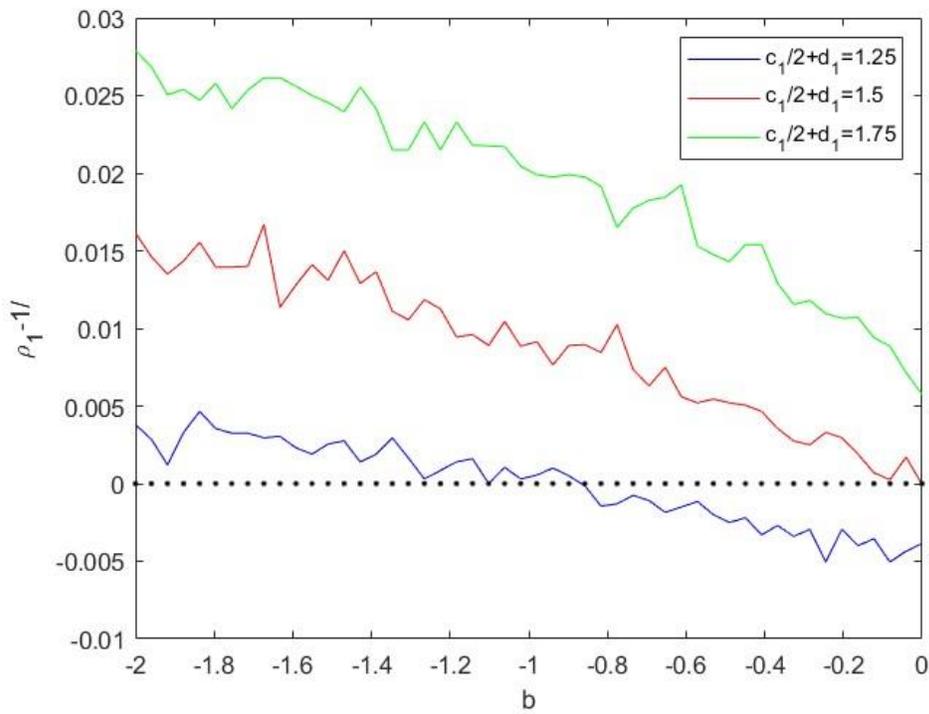

a.

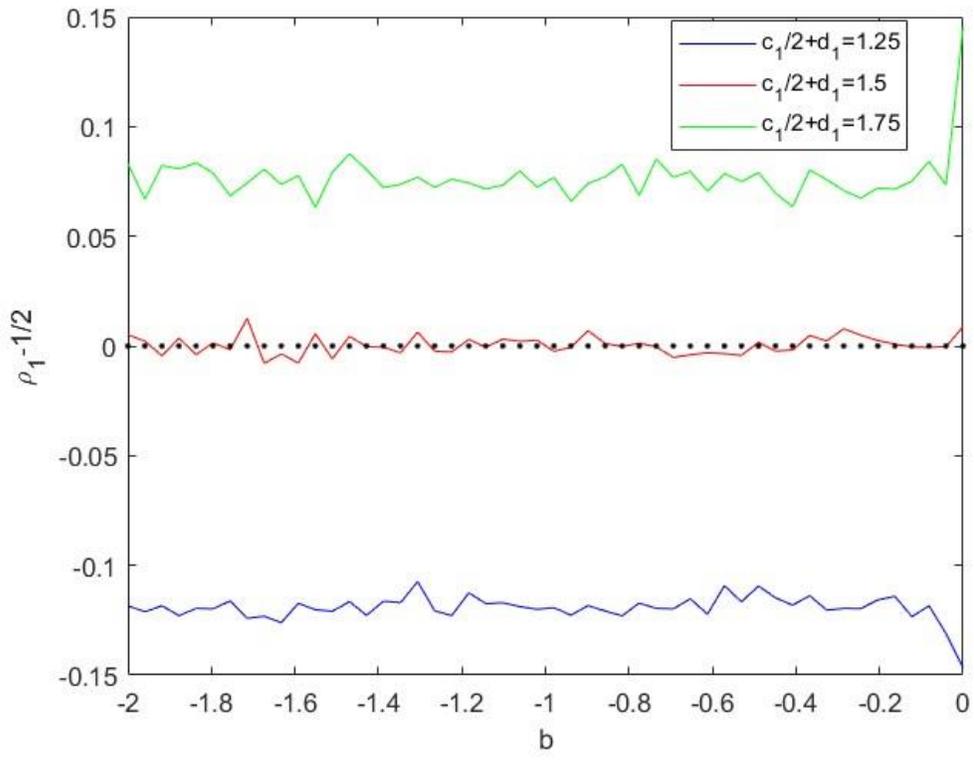

b.

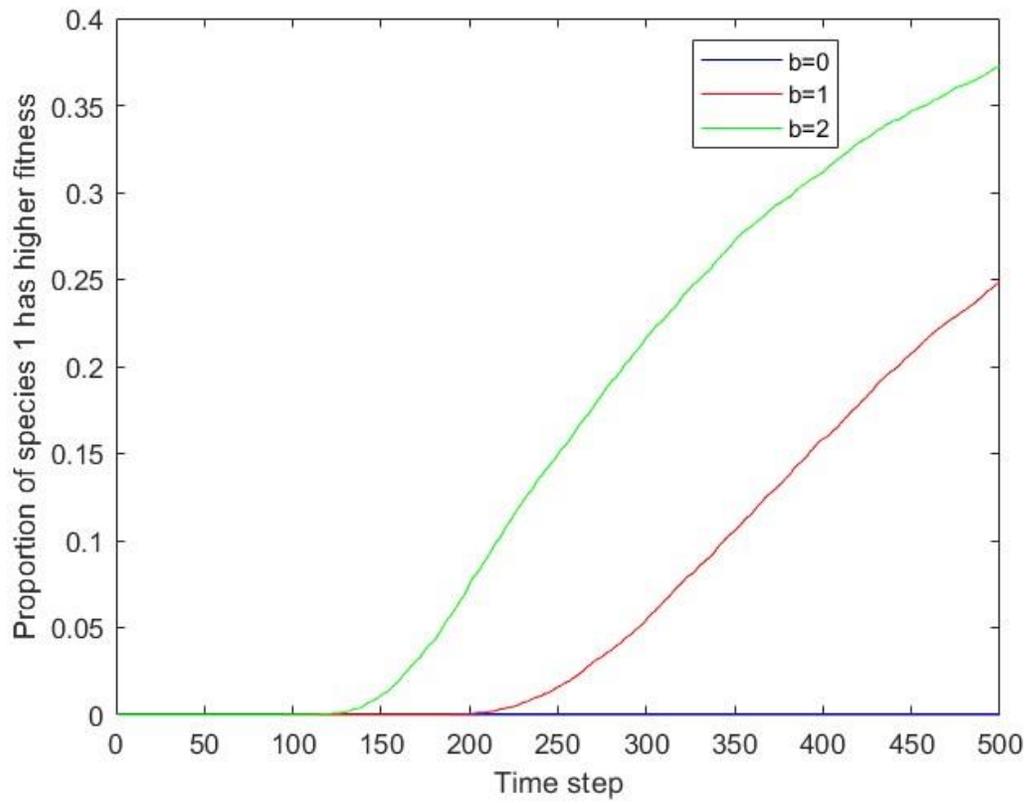

c.

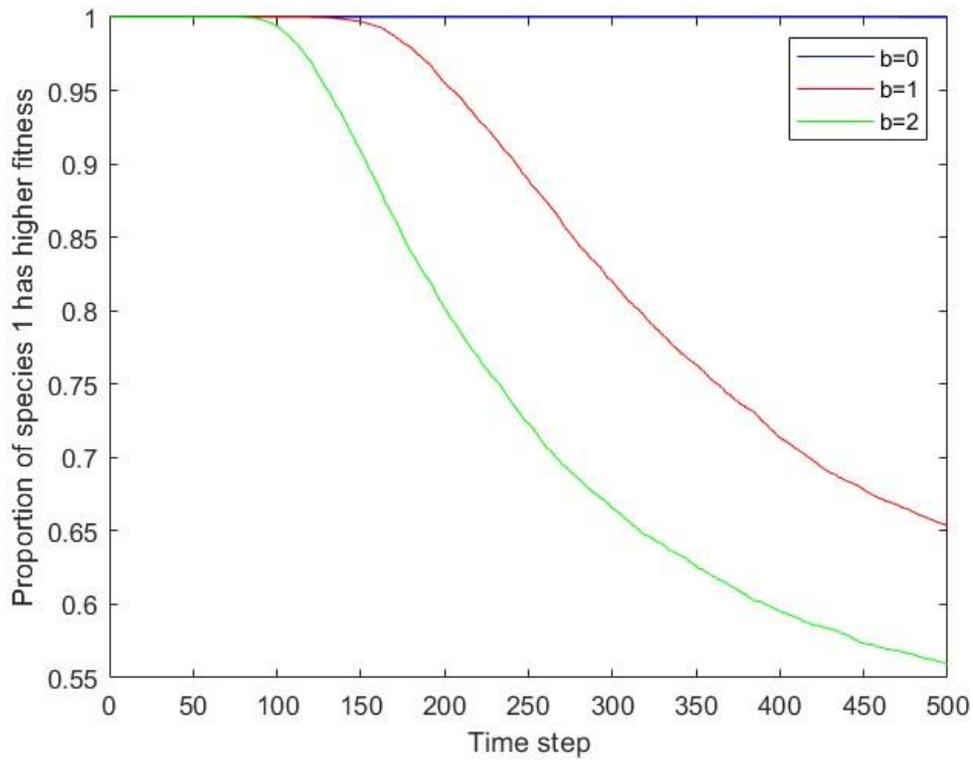

d.

Figure 4. The finite population dynamic is characterized by negative feedback between the environment and the species. **a.** In the invasion scenario, negative feedback increases the possibility of successful invasion of species 2 regardless of whether species 1 is initially dominating, neutral, or inferior. **b.** In a competing environment, negative feedback may weaken the initially dominant species while strengthening the initially inferior species. Negative feedback causes **c.** initially inferior species with $\frac{c_1}{2} + d_1 = 1.25$ to have a better fitness and **d.** dominant species with $\frac{c_1}{2} + d_1 = 1.75$ to lose its advantage in the first 500 time steps. The entire population size is 100, and the selection intensity is 0.05. Each simulation is conducted a total of 10000 times to compute the probabilities $\rho_{\frac{1}{2}}$, $\rho_{\frac{1}{N}}$ and proportion of species 1 has higher fitness than species 2

in first 500 time steps. Other parameter: $\Delta t = 0.001$ and $\frac{c_2}{2} + d_2 = 1.5$.

## 5.Discussion

The evolutionary path often has little influence on the outcome in traditional natural selection theory, especially when there is just one peak on the static fitness landscape, and nature selection is regarded as a separate process that does not take historical accidents into consideration. The fitness landscape, on the other hand, is generally not static but varies, like a fitness seascape where selection changes on micro-evolutionary time scales (Mustonen & Lässig 2009). If we investigate feedback between species and environment, it is a specific instance of fluctuant fitness seascape (Li et al. 2021). We talked about the importance of feedback in two crucial ecological scenarios. To begin, invasion scenarios depict the process by which a single individual from species 1 invades a community of individuals from species 2. Then, competition scenarios depict the process that occurs in areas of overlap in the distribution of two species that have an equal number of individuals at the start.

In the invasion scenarios, natural selection favors species 1 if its fitness is greater than that of species 2 when there is no feedback between environment and species. Positive feedback, on the other hand, makes invasion more difficult, so that even species with high fitness at the start will not be favored by natural selection. Because primordial species have an advantage in terms of individual quantity, the primordial species may be able to make the environment more conducive to itself. Negative feedback, on the other hand, makes large individual number a disadvantage and makes invasion more likely, so that even species with poor fitness may be favored by natural selection. These findings are consistent with experimental findings that nutrient availability can reinforce dominant species' competitive ability and stabilize current vegetation if there is positive feedback between litter effect and nutrient availability, or it can enhance a potential competitor's advantage and lead to vegetation change if there is a negative feedback between litter effect and nutrient availability (Matson 1990; Wedin & Tilman 1990; Aerts 1999).

If there is no feedback in the competitive scenarios, species with better fitness have a greater chance of taking over the whole population. However, both positive and negative feedback might benefit the initially inferior species, increasing its chances of taking over the entire population. The explanation for this is because feedback allows originally inferior species to achieve high fitness. It emphasizes some of the early random events. In other words, in the presence of feedback, when the present state is dependent on the prior path chosen, the evolutionary path is extremely important. This feature means that even initially inferior species have a chance to achieve high fitness without resorting to traditional genetic drift provided they are lucky enough to have a sufficient number of people at the beginning, even if they are of low fitness. Furthermore, the species that dominates the population does not always have the best fitness at the start.